\documentclass[11pt,a4paper]{article}
\pdfoutput=1

\usepackage{jheppub}
\usepackage[utf8]{inputenc}
\usepackage[T1]{fontenc}
\usepackage{ifthen}

\newcommand{\eVq}{\ensuremath{\text{eV}^2}}
\newcommand{\Dmq}{\Delta m^2}
\newcommand{\diag}{\mathop{\mathrm{diag}}}
\newcommand{\Nuc}[2][]{{\ensuremath{\ifthenelse{\equal{#1}{}}{}{\mbox{}^{#1}}\text{#2}}}}

\makeatletter
\DeclareRobustCommand\recite[1]{\begingroup\@fileswfalse\cite{#1}\endgroup}
\makeatother

\graphicspath{{Figures/}}

\allowdisplaybreaks

\title{Solar neutrinos and leptonic spin forces}

\author[a]{Saeed Ansarifard,}
\emailAdd{ansarifard@ipm.ir}
\affiliation[a]{School of physics, Institute for Research in
  Fundamental Sciences (IPM), P.O.\ Box 19395-5531, Tehran, Iran}

\author[b,c,d]{M.~C.~Gonzalez-Garcia,}
\affiliation[b]{Departament de F\'isica Qu\`antica i Astrof\'isica and
  Institut de Ci\`encies del Cosmos, Universitat de Barcelona,
  Diagonal 647, E-08028 Barcelona, Spain}
\affiliation[c]{Instituci\'o Catalana de Recerca i Estudis
  Avan\c{c}ats (ICREA), Pg.\ Lluis Companys 23, E-08010 Barcelona,
  Spain}
\affiliation[d]{C.N.~Yang Institute for Theoretical Physics, Stony
  Brook University, Stony Brook, NY 11794-3840, USA}
\emailAdd{maria.gonzalez-garcia@stonybrook.edu}

\author[e]{Michele Maltoni,}
\emailAdd{michele.maltoni@csic.es}
\affiliation[e]{Instituto de F\'isica Te\'orica (IFT-CFTMAT),
  CSIC-UAM, Calle de Nicol\'as Cabrera 13--15, Campus de Cantoblanco,
  E-28049 Madrid, Spain}

\author[b]{Jo\~ao Paulo Pinheiro,}
\emailAdd{joaopaulo.pinheiro@fqa.ub.edu}

\abstract{We quantify the effects of light spin-zero particles with
  pseudoscalar couplings to leptons and scalar couplings to nucleons
  on the evolution of solar neutrinos.  In this scenario the matter
  potential sourced by the nucleons in the Sun's matter gives rise to
  spin precession of the relativistic neutrino ensemble.  As such the
  effects in the solar observables are different if neutrinos are
  Dirac or Majorana particles.  For Dirac neutrinos the spin-flavour
  precession results into left-handed neutrino to right-handed
  neutrino (\textit{i.e.}, active-sterile) oscillations, while for
  Majorana neutrinos it results into left-handed neutrino to
  right-handed antineutrino (\textit{i.e.}, active-active)
  oscillations.  In both cases this leads to distortions in the solar
  neutrino spectrum which we use to derive constraints on the allowed
  values of the mediator mass and couplings via a global analysis of
  the solar neutrino data.  In addition for Majorana neutrinos
  spin-flavour precession results into a potentially observable flux
  of solar electron antineutrinos at the Earth which we quantify and
  constrain with the existing bounds from Borexino and KamLAND.}

\preprint{IFT-UAM/CSIC-24-67, YITP-SB-2024-08}


\begin{document}

\maketitle

\section{Introduction}

The Standard Model (SM) has been extensively tested at a plethora of
experimental environments and so far no definitive deviation from its
predictions has been observed at the highest energy scales probed,
neither for what concerns the interactions of known
particles~\cite{Pasztor:2019rqu} nor through the appearance of new
heavy states~\cite{Yuan:2020fyf}.  Yet physics beyond the Standard
Model (BSM) is required to address the well-known shortcomings of the
SM.  The obvious conclusion seems to be that there must be a mass gap
between the electroweak and the BSM scales.  There remains, however,
the interesting possibility of new light states that have escaped
detection so far because they are coupled feebly with the SM (see
Ref.~\cite{Antel:2023hkf} for recent review) but can produce tiny yet
observable effects ---~in particular when coherently enhanced.  The
paradigmatic example is light spin-zero axion-like particles: they
entered the spectrum of BSM physics motivated by the strong CP
problem~\cite{Weinberg:1977ma, Wilczek:1977pj, Peccei:1977hh} and they
soon became also a viable solution to the dark matter
puzzle~\cite{Preskill:1982cy, Abbott:1982af, Dine:1982ah}.  Another
notable example is the Majoron, associated with the spontaneous
breaking of the lepton number~\cite{Chikashige:1980ui, Gelmini:1980re}
in models of neutrino mass generation.

In most generality a spin-zero field $\phi$ can couple to the SM model
fermions $\psi$ with a scalar (\textit{i.e.}, spin-independent)
coupling $(\bar\psi\psi\phi)$, or with a pseudoscalar
\emph{spin-dependent} coupling $(\bar\psi\gamma^5\psi\phi)$.  In this
case they generate monopole-monopole forces, as well as
monopole-dipole and dipole-dipole forces~\cite{Moody:1984ba} between
non-relativistic SM fermions.  In particular the dipole interaction is
known to appear as an effective background magnetic field proportional
to the gradient of the $\phi$ field which generates the precession of
the fermion spin.

Recently, monopole-dipole forces have been invoked in the context of
the measurement of the muon anomalous magnetic moment, $(g-2)_\mu$,
which has put in question the consistency with the Standard Model (SM)
theoretical predictions~\cite{Aoyama:2020ynm, Borsanyi:2020mff}.
Concretely, the experimental result in Ref.~\cite{Muong-2:2023cdq} for
the muon anomalous magnetic moment is larger than the data-driven
value predicted in the SM~\cite{Aoyama:2020ynm} by about
$3$--$4\sigma$.  In this context Refs.~\cite{Janish:2020knz,
  Agrawal:2022wjm, Davoudiasl:2022gdg} proposed such monopole-dipole
interaction with the scalar coupling $g_s^N$ to nucleons acting as a
source of the field $\phi$, whose pseudoscalar coupling to the muon,
$g_p^\mu$, will alter its precession frequency and could be
interpreted as a contribution to $(g-2)_\mu$.  For sufficiently light
$\phi$ field, $m_\phi \sim 10^{-14}$~eV, the long range of the force
leads to a coherent enhancement with all atoms in the Earth
contributing to the gradient of the scalar field at Earth's surface
where muon spin experiments are performed.  The required value to
explain the $(g-2)_\mu$ anomaly is $g_s^N g_p^\mu \gtrsim
10^{-30}$~\cite{Agrawal:2022wjm, Davoudiasl:2022gdg}.

New interactions of leptons are known to have also interesting effects
in the neutrino sector because they can affect their flavour evolution
when traveling through large regions of matter, as is the case for
solar neutrinos.  In the SM this leads to the well-known
Mikheev-Smirnov-Wolfenstein (MSW) mechanism~\cite{Wolfenstein:1977ue,
  Mikheev:1986gs} with a well determined potential difference between
the electron neutrino and the other flavours.  New flavour-dependent
interactions can modify the matter potential and consequently alter
the pattern of flavour transitions, thus leaving imprints in the
oscillation data.  In the Standard Model the generated potential is
proportional to the electron density at the neutrino position because
forward elastic scattering takes place in the limit of zero momentum
transfer, so as long as the range of the interaction is shorter than
the scale over which the matter density extends, the effective matter
potential can be obtained in the contact interaction approximation
between the neutrinos and the matter particles.  Conversely, if the
mediator is extremely light the contact interaction approximation is
no longer valid, and the flavour dependent forces between neutrino and
matter particles become long-range.  Neutrino propagation can still be
described in terms of a matter potential, which however is no longer
simply determined by the number density of particles in the medium at
the neutrino position, but it depends instead on the average of the
matter density around it, within a radius of the order of the inverse
of the mass of the interaction mediator~\cite{Grifols:2003gy,
  Joshipura:2003jh, GonzalezGarcia:2006vp, Davoudiasl:2011sz,
  Wise:2018rnb, Smirnov:2019cae, Coloma:2020gfv}.

Moreover if the force couples to the lepton spin, irrespective of its
flavour dependence, it can flip the helicity of the neutrino, a
process usually referred to as spin-flavour precession (SFP) which
takes place, for example, by interaction with the solar magnetic field
in the presence of a flavour-dependent neutrino
magnetic-moment~\cite{Cisneros:1970nq, Okun:1986hi, Okun:1986na,
  Okun:1986uf, Voloshin:1986ty}, and that can be resonantly enhanced
by solar matter~\cite{Akhmedov:1987nc, Lim:1987tk} as well.  An
interesting characteristic of such \emph{spin forces} acting on the
neutrino is that their observable effects are different if neutrinos
are Dirac or Majorana fermions.  For Dirac neutrinos SFP converts a
fraction of the left-handed neutrinos into right-handed neutrinos
which escape detection.  In the context of neutrino oscillations this
corresponds to an active-sterile neutrino oscillation which would
leave an imprint in the solar neutrino data.  If, on the contrary,
neutrinos are Majorana fermions, then SFP results in the conversion
into right-handed antineutrinos which, besides its effect on solar
neutrino data, can lead to a small but potentially observable flux of
solar electron antineutrinos at the Earth.

At present, the global analysis of data from oscillation experiments
provides some of the strongest constraints on the size of interactions
affecting the neutrino flavour evolution either in the
contact~\cite{GonzalezGarcia:2011my, Gonzalez-Garcia:2013usa,
  Esteban:2018ppq, Coloma:2023ixt, Coloma:2022umy} or the
long-range~\cite{Grifols:2003gy, Joshipura:2003jh,
  GonzalezGarcia:2006vp, Coloma:2020gfv} regime, while the
experimental bounds on the solar antineutrino flux provides strong
bounds on the neutrino magnetic-moment due to its associated SFP
effect (for a recent analysis see Ref.~\cite{Akhmedov:2022txm} and
references therein).

With this motivation in this article we study the effect on the
neutrinos flavour oscillations of a pseudoscalar coupling between
neutrinos and a spin-zero particle with scalar couplings to the
nucleons.  In Sec.~\ref{sec:forma} we present the formalism to study
neutrino propagation in the presence of these interactions and the
generated helicity-flipping potential (with details on its derivation
in appendix~\ref{sec:appendix1}).  In Sec.~\ref{sec:results} we derive
bounds on such interaction (as a function of the mediator mass) from
the global analysis of present solar and KamLAND data for either Dirac
or Majorana neutrinos, and from the solar antineutrino flux bounds for
the case of Majorana neutrinos.  Finally, in Sec.~\ref{sec:disc} we
discuss these bounds in the context of the current limits from charged
muon results, and present our conclusions.

\section{Formalism}
\label{sec:forma}

We consider the interactions of a field $\phi$ with scalar couplings
to the nucleons $f \in \{\text{proton}$, $\text{neutron}\}$ and
pseudoscalar couplings to neutrinos:
\begin{equation}
  \label{eq:lagran}
  \mathcal{L}_\phi = \sum_f g_s^f \phi \bar{f}f
  + i\, [g_p^\nu]^{\alpha\beta} \phi\,
  \overline{\nu}_\alpha \gamma^5 \nu_\beta \,,
\end{equation}
where $g_p^\nu$ parametrize the pseudoscalar couplings to the spin of
the neutrinos, which in its most generality is a $3\times 3$ hermitian
matrix in the neutrino flavour space.

Considering solar neutrinos, the interactions in Eq.~\eqref{eq:lagran}
will generate a flavour dependent potential sourced by the nucleons in
the Sun which will affect the flavour evolution of the neutrinos and
can lead to observable signatures.  We present in
Appendix~\ref{sec:appendix1} the derivation of this potential.  In
brief, we find that the interactions in Eq.~\eqref{eq:lagran} generate
a spin-flip potential on a neutrino of energy $E_\nu$ at position
$\vec{x}$ of the form
\begin{equation}
  \label{eq:Vsp}
  V_\text{sp}(\vec{x})
  = -\frac{g_s^f g_p^\nu}{4E_\nu} \mathcal{E}_\perp(\vec{x})
\end{equation}
where $\mathcal{E}_\perp$ is the size of the transverse component
(with respect to the neutrino trajectory) of a vector field
$\vec{\mathcal{E}}$ defined as
\begin{equation}
  \label{eq:nhatdef}
  \vec{\mathcal{E}}(\vec{x})
  \equiv \frac{1}{2\pi} \vec{\nabla}_x
  \int N_f(\vec\rho)\,
    \frac{e^{-m_\phi |\vec\rho - \vec{x}|}}{|\vec\rho - \vec{x}|}\,
    d^3\vec\rho\,
  \equiv \frac{2}{m_\phi^2} \vec\nabla
  \hat{N}_f(\vec{x}, m_\phi) \,.
\end{equation}
In writing Eq.~\eqref{eq:nhatdef} we have introduced the quantity
$\hat{N}_f(\vec{x}, m_\phi)$ which represents the potential-weighted
matter density within a radius $\sim 1/m_\phi$ around the neutrino
location~\cite{Grifols:2003gy, Joshipura:2003jh,
  GonzalezGarcia:2006vp, Davoudiasl:2011sz, Wise:2018rnb}.  Its
normalization factor ensures that $\hat{N}_f(\vec{x}, m_\phi) \to
N_f(\vec{x})$ for $m_\phi \to \infty$, because that is the limit in
which one should recover the contact interaction expectation
(\textit{i.e.}, the neutrino wave packet localized at position
$\vec{x}$ should be directly sensitive to the number density
distribution $N_f(\vec{x})$ of fermion $f$ at such point).

\begin{figure}\centering
  \includegraphics[width=0.85\textwidth]{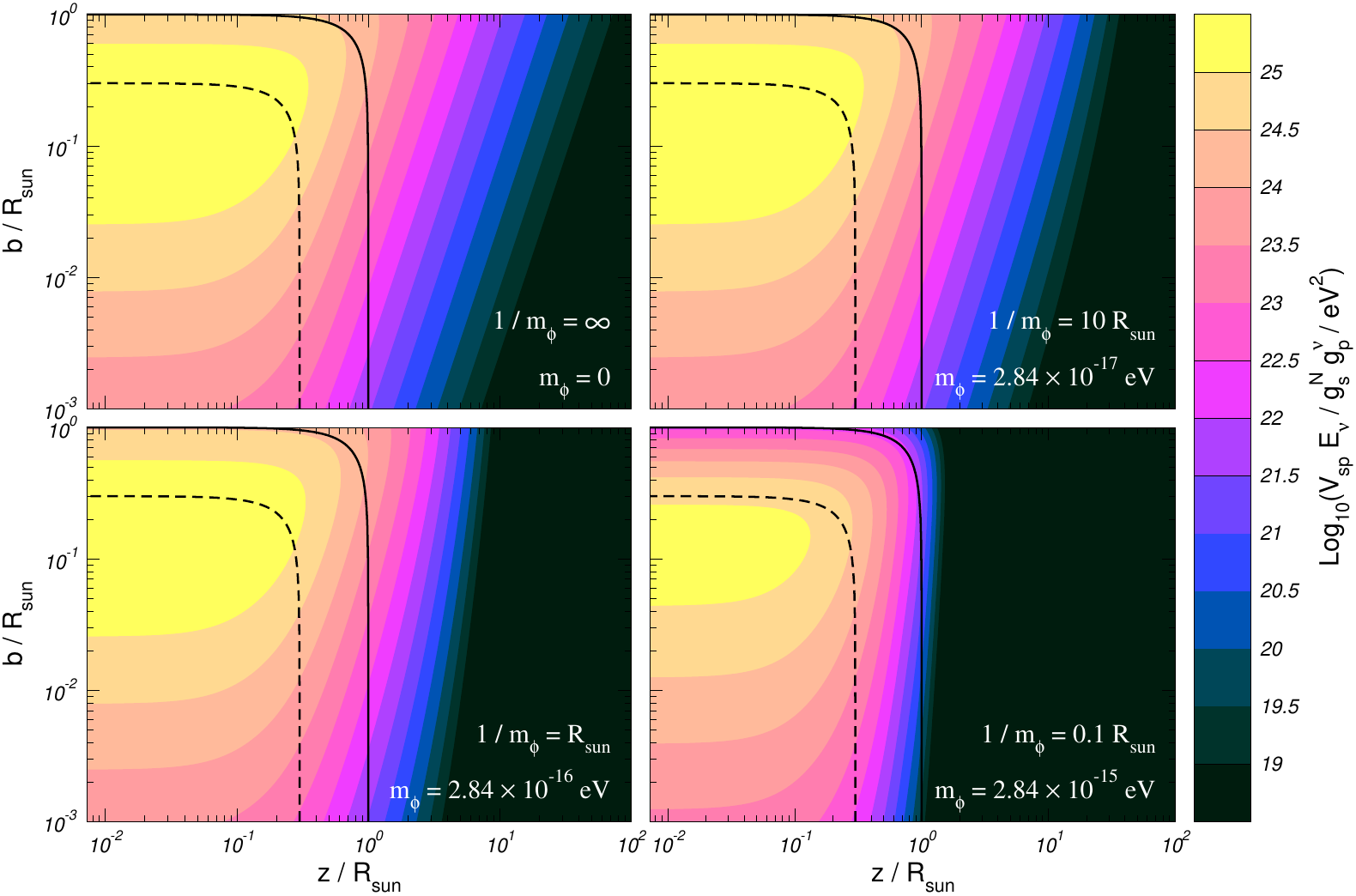}
  \caption{Isocontours of the potential $V_\text{sp}$ sourced by
    nucleons with coupling $g_s^N$ to $\phi$ felt by $\nu$ with
    coupling $g_p^\nu$ to $\phi$ as a function of $b$ (its transverse
    distance to the line connecting the Sun and Earth centers) and $z$
    (the distance along the Sun-Earth direction characterizing its
    trajectory to the Earth) for several values of $m_\phi$.  For
    reference the black full and dashed lines represent the position
    of the edge of the Sun, $\sqrt{b^2+z^2} = R_\odot$, and the outer
    edge of the solar core, $\sqrt{b^2+z^2} = 0.3\, R_\odot$,
    respectively.}
  \label{fig:F3}
\end{figure}

Denoting by $r \equiv |\vec{x} - \vec{x}_\odot|$ the distance from the
center of the Sun $\vec{x}_\odot$ and taking into account the
spherical symmetry of the solar matter distribution $N_f^\odot(r)$, we
see that $\hat{N}_f(\vec{x}, m_\phi)$ simplifies to $\hat{N}_f(r,
m_\phi)$ with:
\begin{equation}
  \hat{N}_f(r, m_\phi)
  = \frac{m_\phi}{2\,r}
  \int_{0}^{R_\odot} \rho\, N_f^\odot(\rho)
  \big[ e^{-m_\phi|\rho-r|} - e^{-m_\phi(\rho+r)} \big] \, d\rho \,.
  \end{equation}
Notice that since $\hat{N}_f(r, m_\phi)$ only depends on the radial
distance to the center of the Sun, the direction of the vector
$\vec{\mathcal{E}}(\vec{x})$ is radial.  As the spin-flip potential is
proportional to the component of $\vec{\mathcal{E}}$ orthogonal to the
neutrino direction, it is clear that it will vanish for trajectories
along the line connecting the solar center and the center of the
Earth.  For neutrinos at a transverse distance $b$ from the line
connecting the Earth and Sun centers, and at a radial distance $r$
from the center of the Sun, the potential is
\begin{equation}
  \label{eq:Vsp2}
  V_\text{sp}(r, b)
  = -\frac{g_s^N g_p^\nu}{2E_\nu\, m_\phi^2}
  \cdot \frac{b}{r} \cdot
  \frac{d\hat{N}_f(r, m_\phi)}{dr} \,.
\end{equation}
In Fig.~\ref{fig:F3} we plot the value of this potential (for
different values of $m_\phi$) at a given position $(b,z)$, where $b$
is the transverse distance to the line connecting the Sun and Earth
centers, and $z=\sqrt{r^2-b^2}$ is the distance along the Sun-Earth
direction (which, as we will see below, is the variable parametrizing
the neutrino trajectory).  For the sake of concreteness we show the
results for $g_s^\text{proton} = g_s^\text{neutron} \equiv g_s^N$.

From the figure we see that, as expected, the potential decreases as
$b\to 0$.  However, it increases very rapidly with $b$ and it reaches
its maximum for values of $b$ well within the neutrino production
region $b\lesssim 0.3\, R_\odot$.  Quantitatively we see that for the
$m_\phi$ considered here the characteristic values of the potential
inside the solar core are comparable with the inverse of the
oscillation length of solar neutrinos with energy $E_\nu$, $E_\nu /
\lambda \sim \Dmq_{21} \sim 10^{-4}~\eVq$ for $g_s^N g_p^\nu \sim
10^{-30}$ within the range of values of the coupling to muons proposed
to account for the $(g-2)_\mu$ anomaly~\cite{Agrawal:2022wjm,
  Davoudiasl:2022gdg}.

To account for the effect of this potential in the solar neutrino
observables we need to solve the evolution equation for the neutrino
ensemble.  Since the potential flips the helicity of the neutrino,
such equation is different for Dirac and Majorana neutrinos.  For a
Dirac neutrino the evolution equation for its flavour components
$\vec\nu\equiv (\nu_e, \nu_\mu, \nu_\tau)^T$ reads
\begin{equation}
  \label{eq:dir}
  i\, \frac{d}{dt}
  \begin{pmatrix}
    \vec\nu_L(t)
    \\
    \vec\nu_R(t)
  \end{pmatrix}
  =
  \begin{pmatrix}
    H^\nu(r) & V_\text{sp}(r, b)
    \\
    V_\text{sp}(r, b)^\dagger & H_\text{vac}
  \end{pmatrix}
  \begin{pmatrix}
    \vec\nu_L(t)
    \\
    \vec\nu_R(t)
  \end{pmatrix},
\end{equation}
where, given the large distance between the Sun and the Earth, we can
assume with high precision that the neutrino travels along a
trajectory which runs parallel to the line connecting the Sun and
Earth centers, so that its distance $b$ from such line remains
constant during propagation.  Denoting by $z$ the longitudinal
position of the neutrino along its trajectory, we have that at a given
time $t$ the distance from the neutrino location to the center of the
Sun is given by $r=\sqrt{b^2+(z-z_0)^2}$ with $z = z_0 + c t$.

For a Majorana neutrino the evolution equation is
\begin{equation}
  \label{eq:maj}
  i\, \frac{d}{dt}
  \begin{pmatrix}
    \vec\nu_L(t)
    \\
    \vec{\overline\nu}_R(t)
  \end{pmatrix}
  =
  \begin{pmatrix}
    H^\nu(r) & V_\text{sp}(r, b)
    \\
    V_\text{sp}(r, b)^\dagger & H^{\overline{\nu}}(r)
  \end{pmatrix}
  \begin{pmatrix}
    \vec\nu_L(t)
    \\
    \vec{\overline\nu}_R(t)
  \end{pmatrix},
\end{equation}
where
\begin{equation}
  H^\nu(r) = H_\text{vac} + H_\text{mat} (r)
  \quad\text{and}\quad
  H^{\overline{\nu}}(r) = [H_\text{vac} - H_\text{mat}(r)]^* \,.
\end{equation}
Here $H_\text{vac}$ is the vacuum part which in the flavour basis
$(\nu_e, \nu_\mu, \nu_\tau)$ reads
\begin{equation}
  \label{eq:Hvac}
  H_\text{vac} = U_\text{vac} D_\text{vac} U_\text{vac}^\dagger
  \quad\text{with}\quad
  D_\text{vac} = \frac{1}{2E_\nu} \diag(0, \Dmq_{21}, \Dmq_{31})
\end{equation}
where $U_\text{vac}$ denotes the three-lepton mixing matrix in
vacuum~\cite{Pontecorvo:1967fh, Maki:1962mu, Kobayashi:1973fv}.
Following the convention of Ref.~\cite{Coloma:2016gei}, we define
$U_\text{vac} = R_{23}(\theta_{23}) R_{13}(\theta_{13})
\tilde{R}_{12}(\theta_{12}, \delta_\text{CP})$, where
$R_{ij}(\theta_{ij})$ is a rotation of angle $\theta_{ij}$ in the $ij$
plane and $\tilde{R}_{12}(\theta_{12}, \delta_\text{CP})$ is a complex
rotation by angle $\theta_{12}$ and phase $\delta_\text{CP}$.
$H_\text{mat}(r)$ is the matter potential due to the SM weak
interactions, which in the flavour basis takes the diagonal form
\begin{equation}
  \label{eq:hmsw}
  H_\text{mat}(r) = \frac{G_F}{\sqrt{2}}
  \diag\Big[ 2N_e(r) -N_n(r),\, -N_n(r),\, -N_n(r) \Big]
\end{equation}
and $N_e(r)$ and $N_n(r)$ are the electron and neutron number
density at distance $r$ from the solar center.

We numerically solve the evolution equation from the production point
of the ${\nu_e}_L$ in the Sun to the Earth.  With our choice of
variables we can characterize the production point by its radial
distance $r_0$ from the center of the Sun, its transverse distance $b$
from the line connecting the Sun-Earth centers, and the angle $\phi$
in that transverse plane, and we are left with two discrete
possibilities $\vec{x}_0^\pm = (\pm \sqrt{r_0^2-b^2}, b\cos\phi_0,
b\sin\phi_0)$ where positive (negative) sign corresponds to neutrinos
produced in the hemisphere of the Sun which is closer (further) from
the Earth.  In this way we obtain the oscillation probability into
$\nu_\alpha$ (here denoting generically either a left-handed neutrino
state, or a right-handed neutrino for Dirac and a right-handed
antineutrino for Majorana), $P_{e\alpha}(\vec{x}_0^\pm, E_\nu)$.  In
the Standard Solar Models the probability distribution of the neutrino
production point only depends on the radial distance to the center of
the Sun, $r_0$ (with different distributions for the different
neutrino production reactions) so it is convenient to define
\begin{equation}
  P_{e\alpha}(r_0, E_\nu)
  \equiv
  \frac{1}{4\pi}\int d\Omega\, P_{e\alpha}(\vec{x}_0, E_\nu)
  =\frac{1}{4\pi r_0}\int_0^{r_0} db\, \frac{b}{\sqrt{r_0^2-b^2}}
    \int_0^{2\pi} d\phi\, \sum_{\pm}P_{e\alpha}(\vec{x}_0^\pm, E_\nu) \,.
\end{equation}
As mentioned above each of the eight reactions producing solar
neutrinos ---~labeled by $i=1\dots 8$ for \Nuc{pp}, \Nuc[7]{Be},
\Nuc{pep}, \Nuc[13]{N}, \Nuc[15]{O}, \Nuc[17]{F}, \Nuc[8]{B}, and
\Nuc{hep}~--- generates neutrinos with characteristic distributions
$\mathcal{R}_i(r_0)$ (normalized to one) which are energy independent.
Thus we can obtain the mean survival probability for a neutrino of
energy $E_\nu$ produced in reaction $i$ as
\begin{equation}
  P^{i}_{e\alpha}(E_\nu)
  = \int_0^{R_{\odot}} dr_0 \, \mathcal{R}_i(r_0)\, P_{e\alpha}(r_0, E_\nu)
\end{equation}
and with those obtain the predictions for the solar observables in our
analysis.

\begin{figure}\centering
  \includegraphics[width=\textwidth]{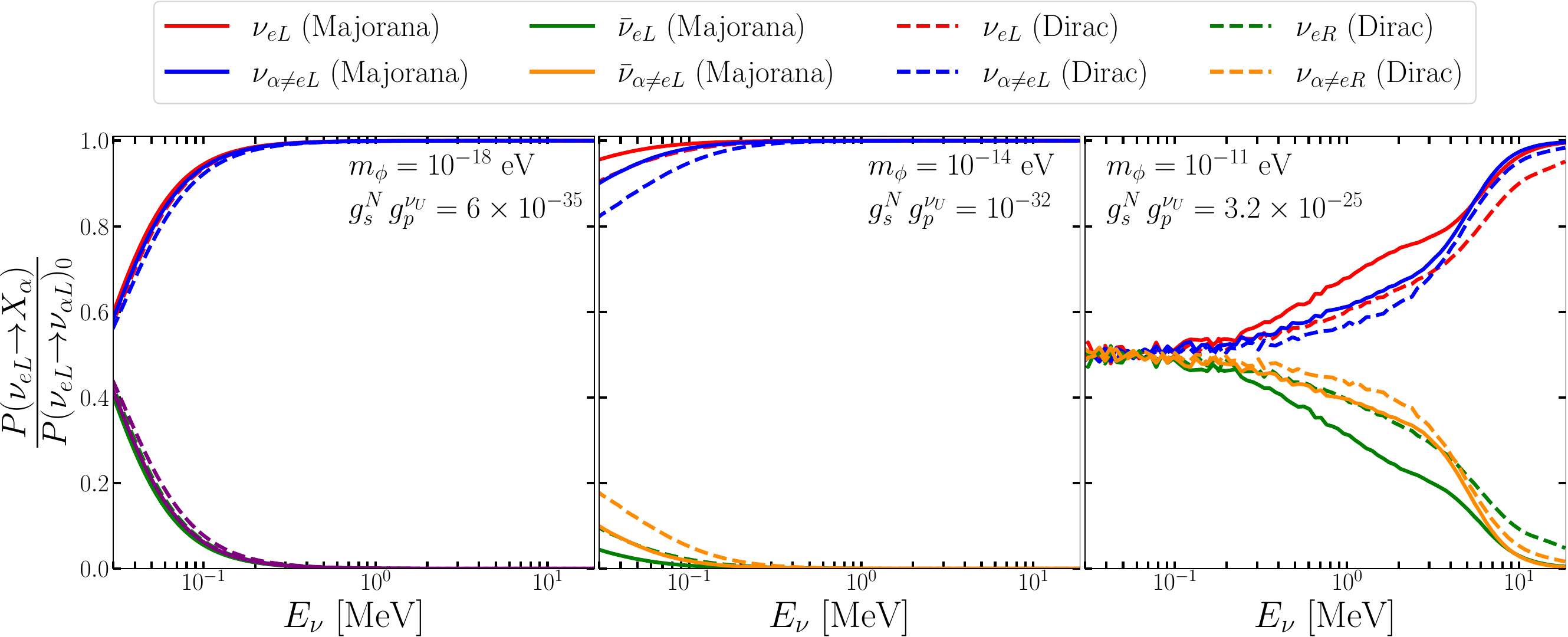}
  \caption{Relevant oscillation probabilities at the Earth surface of
    solar $\nu_{eL}$ into left and right neutrinos (antineutrinos) of
    flavour $\alpha$ divided by $P(\nu_{e_L}\to \nu_{\alpha L})_0$
    (the probability for zero coupling to the $\phi$ field) as a
    function of the neutrino energy for different values of model
    parameters as labeled in the figure.  Oscillation parameters have
    been fixed to $\Dmq_{21} = 7.4\times 10^{-5}~\eVq$,
    $\sin^2\theta_{12} = 0.303$, $\sin^2\theta_{13} = 0.02203$,
    $\Dmq_{31} = 2.5\times 10^{-3}~\eVq$, $\sin^2\theta_{23} = 0.5$,
    and $\delta_\text{CP} = 0$.}
  \label{fig:probs}
\end{figure}

In what follows, for the sake of concreteness, we are going to focus
on four cases: three in which $\phi$ couples to a specific flavour
$\gamma$ and a fourth in which the coupling is universal in flavour
space
\begin{equation}
  \label{eq:nucoup}
  [g_p^\nu]^{\alpha\beta}
  = g_p^{\nu_\gamma} \, \delta^{\alpha\gamma}\delta^{\beta\gamma}
  \enspace\text{for}\enspace
  \gamma \in \{e, \mu, \tau \} \,,
  \quad\text{and}\quad
  [g_p^\nu]^{\alpha\beta}
  = g_p^{\nu_U} \, \delta^{\alpha\beta}
  \enspace\text{for universal}.
\end{equation}
As illustration we plot in Fig.~\ref{fig:probs} the relevant
oscillation probabilities for Dirac and Majorana cases as a function
of the neutrino energy.  The figure is shown for the case of coupling
to $\nu_U$ but the behaviour is qualitatively the same for couplings
to $\nu_e$, $\nu_\mu$ or $\nu_\tau$.\footnote{For concreteness the
probabilities shown here have been generated integrating over the
production point distribution $\mathcal{R}_\text{\Nuc[8]{B}}(r_0)$,
but the corresponding plots obtained with the distribution
probabilities for the other solar fluxes are very similar.}  We assume
fixed oscillation parameters $\Dmq_{21} = 7.4\times 10^{-5}~\eVq$,
$\sin^2\theta_{12} = 0.303$, $\sin^2\theta_{13} = 0.02203$, $\Dmq_{31}
= 2.5\times 10^{-3}~\eVq$, $\sin^2\theta_{23}=0.5$, and
$\delta_\text{CP}=0$, and display different values for the two model
parameters ($m_{\phi}$, $g_s^N g_p^{\nu_U}$).  In the left (central)
panel $m_\phi$ is within the infinite (contact) range interaction and
the coupling constants are close to what will be the maximum allowed
value, while the right panel corresponds to parameters which will be
well within the region ruled out by our analysis, see
Fig.~\ref{fig:limitsnu}.  As seen in the figure the relevant
probabilities are not very different for Dirac and Majorana neutrinos.
Most importantly the figure shows that for moderate values of the
model parameters the new interaction has largest effect on the
disappearance of $\nu_{eL}$ and the appearance of $\nu_{\alpha\neq
  e,L}$ and $\nu_{\alpha R}$ ($\overline{\nu}_{\alpha R}$) for Dirac
(Majorana) neutrinos with the lowest energies.  This is expected as
the potential in Eq.~\eqref{eq:Vsp} is inversely proportional to the
neutrino energy, as characteristic of interactions with spin-zero
mediators~\cite{GonzalezGarcia:2006vp}.  This is at a difference with
the helicity-conserving MSW potential of Eq.~\eqref{eq:hmsw} (or with
any vector interaction in general~\cite{Coloma:2020gfv}), and with the
helicity-flip potential generated by a neutrino magnetic
moment~\cite{Akhmedov:1987nc, Lim:1987tk} which are independent of the
neutrino energy.  We also see that for these cases the relative effect
on disappearance of $\nu_{eL}$ with $\sim
\mathcal{O}(\text{10-100})$~KeV energy is larger than the
$\overline{\nu}_{eR}$ appearance at MeV.  These two facts are of
relevance in the derivation of the bounds from the analysis of the
solar neutrino and antineutrino data as we describe next.

\section{Results}
\label{sec:results}

\subsection{Bounds from global analysis of solar neutrino oscillation data}

We first perform a global fit to solar neutrino oscillation data in
the framework of three massive neutrinos including the new
neutrino-matter interactions generated by the Lagrangian in
Eq.~\eqref{eq:lagran} with the four choices of pseudoscalar couplings
to neutrinos in Eq.~\eqref{eq:nucoup}.  For the detailed description
of the methodology we refer to our latest published global analysis
NuFIT-5.0~\cite{Esteban:2020cvm} to which we have included the data
additions in NuFIT-5.3~\cite{nufit-5.3}; some technical aspects of our
treatment of neutrino propagation in the solar matter can be found in
Sec.~2.4 of Ref.~\cite{Maltoni:2023cpv}.  For solar neutrinos the
analysis includes the total rates from the radiochemical experiments
Chlorine~\cite{Cleveland:1998nv}, Gallex/GNO~\cite{Kaether:2010ag},
and SAGE~\cite{Abdurashitov:2009tn}, the spectral and zenith data from
the four phases of Super-Kamiokande (SK) in Refs.~\cite{Hosaka:2005um,
  Cravens:2008aa, Abe:2010hy} including the latest SK4 2970-day energy
and day/night spectrum~\cite{Super-Kamiokande:2023jbt}, the results of
the three phases of SNO in the form of the day-night spectrum data of
SNO-I~\cite{Aharmim:2007nv} and SNO-II~\cite{Aharmim:2005gt} and the
three total rates of SNO-III~\cite{Aharmim:2008kc}, and the spectra
from Borexino Phase-I~\cite{Bellini:2011rx, Bellini:2008mr} (BX1),
Phase-II~\cite{Borexino:2017rsf} (BX2), and
Phase-III~\cite{BOREXINO:2022abl} (BX3).  In the framework of
three-neutrino mixing the oscillation probabilities for solar
neutrinos dominantly depend on $\theta_{12}$ and $\Dmq_{21}$, for
which relevant constraints arise from the analysis of the KamLAND
reactor data, hence we include in our fit the separate DS1, DS2, DS3
spectra from KamLAND~\cite{Gando:2013nba}.  Finally, let us mention
that, in principle, the Earth matter nucleons also generate a
spin-flavour precession potential which could affect the evolution of
the KamLAND reactor antineutrinos as well as the day and night solar
neutrino variations.  Quantitatively, however, for the range of model
parameters considered here this effect is negligible compared to that
induced by the Sun, and it can therefore be safely neglected.

In what respect the relevant parameter space, Eqs.~\eqref{eq:dir}
and~\eqref{eq:maj} are a set of six coupled linear differential
equations whose solution depends on the two model parameters
($m_{\phi}$, $g_s^N g_p^{\nu_a}$) (for $a=e, \mu, \tau$ or $U$) and
the six oscillation parameters, $\theta_{12}$, $\theta_{13}$,
$\theta_{23}$, $\Dmq_{21}$, $\Dmq_{32}$, and $\delta_\text{CP}$.
However, as it is well-know, the present determination of $\Dmq_{32}$
---~derived dominantly from atmospheric, long-baseline accelerator,
and medium-baseline reactor neutrino experiments~--- implies that the
solar neutrino oscillations driven by $\Dmq_{32} / E_\nu$ are averaged
out.  Thus in our numerical evaluations we fix it to its present
best-fit value, but any other value would yield the same result as
long as we remain in the regime of averaged $\Dmq_{32}$
oscillations.\footnote{In fact this would allow for a reduction of the
evolution equations to an effective $4\times 4$ system.  However from
the computational point of view the reduction from a $6\times 6$ to a
$4\times 4$ set of equations is not substantial, so in our
calculations we stick to the complete $6\times 6$ system.}  In the
standard $3\nu$ oscillation scenario $\nu_\mu$ and $\nu_\tau$ are
totally indistinguishable at the energy scale of solar and reactor
experiments, hence also the dependence on $\theta_{23}$ and
$\delta_\text{CP}$ drops out in the evaluation of the solar
observables.  Including the flavour dependent scalar-pseudoscalar
potential reintroduces a mild dependence on $\theta_{23}$ and
$\delta_\text{CP}$ (except for the flavour-universal case $a=U$), but
quantitatively the phenomenology of solar neutrino data is still
dominated by $\Dmq_{21}$ and $\theta_{21}$.  The dependence on
$\theta_{13}$ is much weaker than its present precision from
medium-baseline reactor experiments (in particular from
Daya-Bay~\cite{DayaBay:2022orm}), while the impact of the CP phase is
very marginal.  So we can safely fix $\theta_{13}$ to its current
best-fit value from NuFIT-5.3 $\sin^2\theta_{13} = 0.02203$ and set
$\delta_\text{CP}=0$.  In what respects $\theta_{23}$, for coupling to
specific flavours $a=e,\mu,\tau$ we marginalize it within the allowed
range from the global oscillation analysis by adding a prior for
$\sin^2\theta_{23}$ from NuFIT-5.3~\cite{nufit-5.3} to the $\chi^2$
function of the solar and KamLAND experiments.\footnote{Our results,
however, show that the difference in the bounds obtained marginalizing
over $\theta_{23}$ with this prior or fixing $\theta_{23}$ to its
best-fit value is very small.}  Altogether the parameter space to be
explored is five-dimensional.

We perform the analysis for both Dirac and Majorana neutrinos.
Besides the different $\nu_{eL}$ oscillation probabilities resulting
from the evolution equations Eqs.~\eqref{eq:dir} and~\eqref{eq:maj},
there are also differences in the evaluation of the neutrino-electron
elastic scattering event rates in Borexino, SNO and SK: while for the
Dirac case the produced $\nu_{\alpha R}$ does not interact in the
detector, for Majorana neutrinos the $\overline{\nu}_{\alpha R}$ do
interact (albeit with a different cross section than the corresponding
$\nu_{\alpha L}$ neutrinos) and such interactions must be included in
the evaluation of the event rates.  Furthermore the CC event rate at
SNO can also be different for Dirac and Majorana because the
$\overline{\nu}_{e R}$ interaction will produce a positron and two
neutrons which may or may not be discriminated.  Lacking the details
for a proper estimation, we have performed the analysis under two
extreme hypothesis: either assuming that the $e^+$ is recognized and
the whole event is rejected, or that the positron is misidentified as
an electron and it contributes as a CC event \emph{plus} two
additional NC events because of the generated neutrons.  We have
verified that the final bounds obtained under the two assumptions are
the same.

\begin{figure}\centering
   \includegraphics[width=\textwidth]{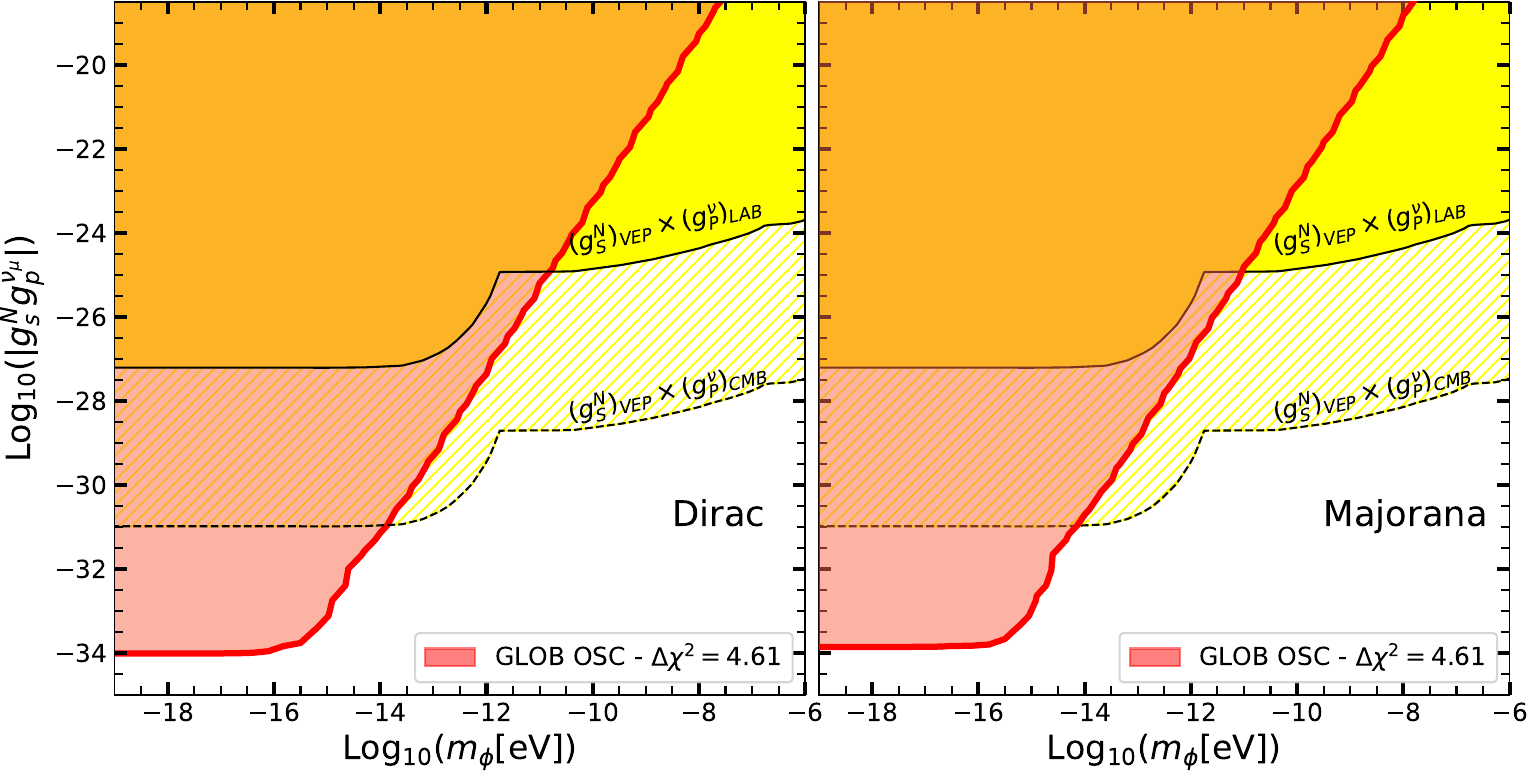}
   \caption{Excluded region at 90\% CL in the ($m_{\phi}$, $g_s^N
     g_p^{\nu_U}$) plane by the global analysis of solar neutrino data
     after marginalization over the relevant oscillation parameters
     ($\Dmq_{21}$, $\theta_{12}$) while keeping the other oscillation
     parameters fixed as described in the text.  Left (right) panel
     corresponds to Dirac (Majorana) neutrinos.  For comparison we
     show as yellow region the bounds on the product of couplings from
     the product of the bound on $g_s^N$ from violation of weak
     equivalent principle (VEP) and the bounds on $g_p^\nu$ from
     either kinematic and rare effects in weak decays in laboratory
     experiments (LAB) or cosmology bounds from cosmic microwave
     background (CMB) (see Sec.~\ref{sec:disc}).}
   \label{fig:limitsnu}
\end{figure}

Altogether we find that the inclusion of the new interaction does not
lead to any significant improvement on the description of the data,
and in fact the overall best-fit $\chi^2$ in the five-dimensional
parameter space is the same as in the standard $3\nu$ oscillation
scenario.  Consequently the statistical analysis results into bounds
on the allowed range of the scalar-pseudoscalar model parameters.

We show in Fig.~\ref{fig:limitsnu} the excluded region on the
parameter space for the case of flavour-universal coupling
($m_{\phi}$, $g_s^N g_p^{\nu_U}$) derived from the global analysis of
solar neutrino and KamLAND data for Dirac and Majorana cases after
marginalization over the relevant oscillation parameters ($\Dmq_{21}$,
$\theta_{12}$) while keeping the other oscillation parameters fixed to
their best-fit values as discussed above.  We show the regions at 90\%
CL (2~dof, two-sided).  Notice that when reporting the bounds on the
model parameters one has a choice on the statistical criteria to
derive the constraints.  The reason is that the experiment is in fact
only sensitive to $|g_s^N g_p^{\nu_U}|$: since the potential is
helicity-flipping its effect does not interfere with the
helicity-conserving vacuum oscillation and MSW matter potential, so
the probabilities only depend on the absolute value of the couplings
and on the square of the mediator mass.  Therefore, accounting for the
physical boundary, it is possible to report the limit on the absolute
value of the coupling (and mass) as a \emph{one-sided} limit, which at
90\% CL for 1~dof (2~dof) corresponds to $\Delta\chi^2 = 1.64$
($3.22$).  Conversely, if this restriction is not imposed, the result
obtained is what is denoted as a \emph{two-sided} limit, which at
90\%CL for 1~dof (2~dof) corresponds to $\Delta\chi^2 = 2.71$ ($4.61$)
and results into weaker constraints.  In what follows we list the
bounds obtained with the least constraining two-sided criterion.

As seen in Fig.~\ref{fig:limitsnu} the results are quantitatively
similar for Dirac or Majorana neutrinos.  In the figure we observe a
change in the slope of the exclusion region for masses $m_\phi\sim
10^{-15}$~eV.  For smaller mediator masses, the interaction length is
larger than the Sun radius and the corresponding potential becomes
saturated.  Conversely for $m_\phi\gtrsim 10^{-14}$~eV the interaction
range is short enough for the contact interaction approximation to
hold, in which case the analysis only depends on the combination
$|g_s^N g_p^{\nu_U}| \,\big/\, m_\phi^2$ and the region boundary
becomes a straight line of slope two in the log-log plane.

\begin{figure}\centering
  \includegraphics[width=0.95\textwidth]{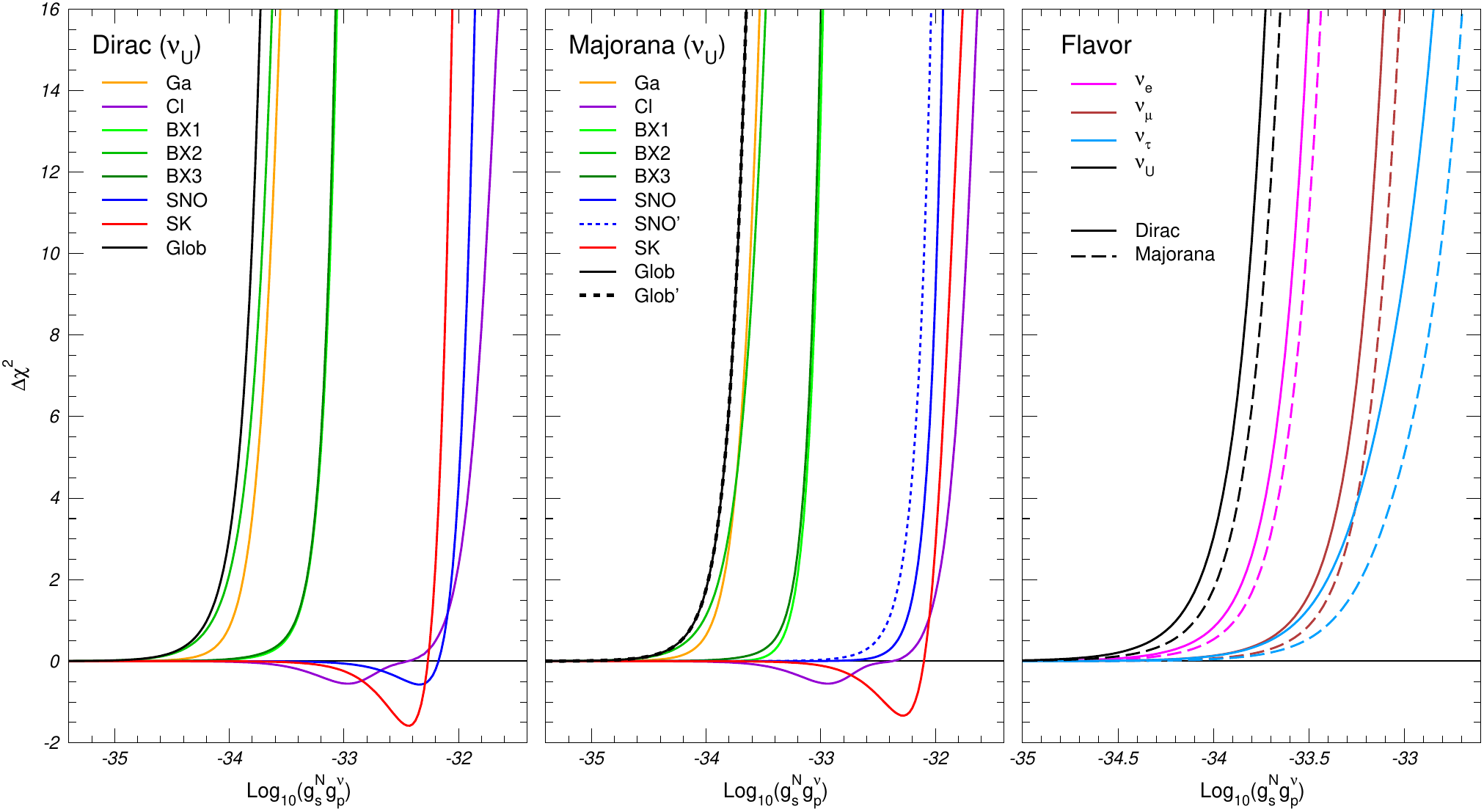}
  \caption{$\Delta\chi^2 \equiv \chi^2(g_s^N g_p^{\nu_U}) -
    \chi^2(g_s^N g_p^{\nu_U} = 0)$ as a function of $g_s^N
    g_p^{\nu_U}$ for $m_\phi\ll 10^{-15}$~eV fixing all oscillation
    parameters (in particular $\Dmq_{21}$, $\theta_{12}$ and
    $\theta_{23}$) to their best-fit values.  The left (central) panel
    show $\Delta\chi^2$ for each of the individual solar neutrino
    experiments for Dirac (Majorana) neutrinos and for the case of
    coupling to $\nu_U$.  In the central panel SNO and SNO' label the
    two variants of the SNO CC analysis for Majorana neutrinos (see
    text for details).  The right panel show the global $\Delta\chi^2$
    for four choices of the flavour dependence of the
    pseudoscalar-neutrino couplings; in particular, the black lines
    (coupling to $\nu_U$) for Dirac and Majorana cases coincide with
    the corresponding ``Glob'' ones in left and central panels,
    respectively.}
  \label{fig:solexp}
\end{figure}

To illustrate the relevance of the different solar neutrino
experiments on these results we plot in the left (central) panels in
Fig.~\ref{fig:solexp} the value of $\Delta\chi^2 \equiv \chi^2(g_s^N
g_p^{\nu_U}) - \chi^2(g_s^N g_p^{\nu_U} = 0)$ as a function of $g_s^N
g_p^{\nu_U}$ for the Dirac (Majorana) case in the effective
infinite-range interaction limit for each of the individual solar
neutrino experiments, fixing all oscillation parameters (in particular
$\Dmq_{21}$, $\theta_{12}$ and $\theta_{23}$) to their best-fit
values.  We see from the figure that the constraints are driven by the
experiments sensitive to the lowest energy part of the solar neutrino
spectrum (\textit{i.e.}, to the \Nuc{pp} flux), which correspond to
the spectral data of phase-II of Borexino (BX2 in the figure) and the
event rates in Gallium experiments (Ga).  Conversely, those most
sensitive to the higher energy part of the solar spectrum
(\textit{i.e.}, the \Nuc[8]{B} flux), such as the spectral information
of Super-Kamiokande and SNO and the total rate in Chlorine (Cl), yield
weaker sensitivity to the new interaction.  As a consequence the
combined constraints for Majorana neutrinos are the same for both
variants of the SNO CC analysis (labeled as SNO and SNO' in the
figure).  Furthermore, since the oscillation parameters $\Dmq_{21}$
and $\theta_{12}$ are dominantly determined by KamLAND reactor data as
well as SNO and SK solar neutrino data, their determination is very
little affected by the inclusion of the new interaction, and the
allowed ranges that we found in this five-parameter analysis are
exactly the same as in the standard $3\nu$ oscillation.

The right panel illustrates the dependence of the combined bounds on
the assumption for the neutrino flavour of the pseudoscalar coupling
for the four choices in Eq.~\eqref{eq:nucoup}.  As seen the variation
in the bounds is below $\mathcal{O}(10)$.  As expected the difference
is larger between coupling to $\nu_e$ and to $\nu_\mu$ (or $\nu_\tau$)
because the coupling to $\nu_e$ has larger effect on $P_{ee}$ which
gives the dominant contribution to the event rates in solar neutrino
experiments.  We notice in passing that the small variation between
the bounds for couplings to $\nu_\mu$ and $\nu_\tau$ is dominantly due
to the non-zero value of $\theta_{13}$.\footnote{In fact, it can be
shown that in the limit $\theta_{13} = 0$ the $\chi^2$ function
depends on $g_s^N g_p^{\nu_\mu}$, $g_s^N g_p^{\nu_\tau}$ and
$\theta_{23}$ only through the effective combination $g_s^N
g_p^{\nu_\mu} \cos^2\theta_{23} + g_s^N g_p^{\nu_\tau}
\sin^2\theta_{23}$.  In the general case $\theta_{13} \ne 0$ such
relation becomes only approximate.}

Finally we list in table~\ref{tab:bounds} the 90\% CL (1~dof) bounds
obtained from the global analysis of solar data in the two asymptotic
regimes and for the four choices of flavour dependence after
marginalization over the three relevant oscillation parameters
$\Dmq_{21}$, $\theta_{12}$ and $\theta_{23}$.

\begin{table}\centering
  \begin{tabular}{|c|c||c|c||c|}
    \hline
    \multicolumn{2}{|c||}{}
    & \multicolumn{2}{c||}{Global Solar $\nu$} &
    Solar $\overline{\nu}_e$
    \\
    \cline{3-5}
    \multicolumn{2}{|c||}{}
    & Dirac & Majorana & Majorana
    \\
    \hline
    & $a=e$
    & $1.5\times 10^{-34}$
    & $1.7\times 10^{-34}$
    & $1.1\times 10^{-33}$
    \\
    $\big| g_s^N g_p^{\nu_a} \big|$
    & $a=\mu$
    & $3.6\times 10^{-34}$
    & $4.3\times 10^{-34}$
    & $7.6\times 10^{-34}$
    \\
    for $m_\phi< 10^{-16}$~eV
    & $a=\tau$
    & $3.6\times 10^{-34}$
    & $6.3\times 10^{-34}$
    & $7.9\times 10^{-34}$
    \\
    & $a=U$
    & $7.6\times 10^{-35}$
    & $9.6\times 10^{-35}$
    & $3.2\times 10^{-34}$
    \\\hline
    & $a=e$
    & $1.3\times 10^{-31}$
    & $2.5\times 10^{-31}$
    & $1.9\times 10^{-31}$
    \\
    $\frac{\left|g_s^N g_p^{\nu_a}\right|}{m_\phi \,/\, 10^{-14}~\eVq}$
    & $a=\mu$
    & $2.0\times 10^{-31}$
    & $5.0\times 10^{-31}$
    & $2.5\times 10^{-31}$
    \\
    for $m_\phi> 10^{-14}$~eV
    & $a=\tau$
    & $2.0\times 10^{-31}$
    & $7.9\times 10^{-31}$
    & $2.5\times 10^{-31}$
    \\
    & $a=U$
    & $6.3\times 10^{-32}$
    & $1.3\times 10^{-31}$
    & $7.0\times 10^{-32}$
    \\
    \hline
  \end{tabular}
  \caption{Bounds from the different analysis in the two asymptotic
    regimes, effective infinite interaction range and contact range
    respectively, and for the four choices of flavour dependence.  For
    the global solar analysis the bounds shown are obtained at 90\% CL
    (1~dof, two-sided), $\Delta\chi^2 = 2.71$.  The bounds from the
    antineutrino flux constraint are obtained from the 90\% CL
    experimental constraints in Eq.~\eqref{eq:panubound}.  See text
    for details.}
  \label{tab:bounds}
\end{table}

\subsection{Bounds from solar antineutrino constraints}

\begin{figure}\centering
  \raisebox{0.2mm}{\includegraphics[width=0.48\textwidth]{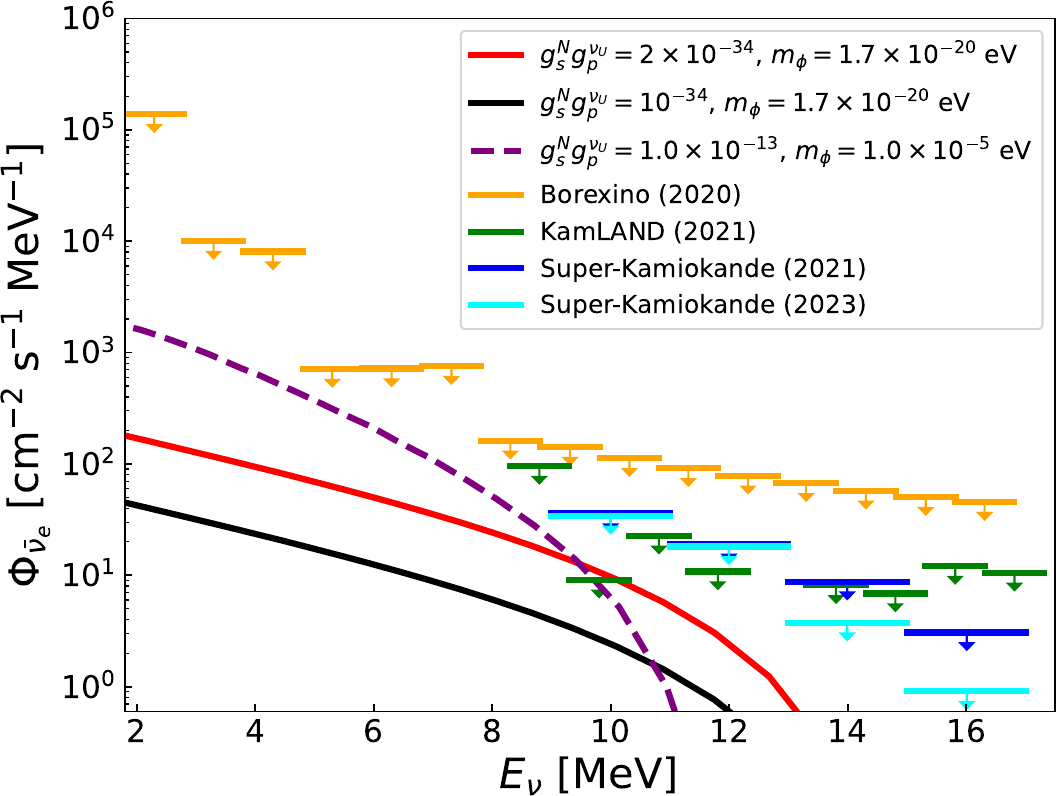}}
  \hfill\includegraphics[width=0.50\textwidth]{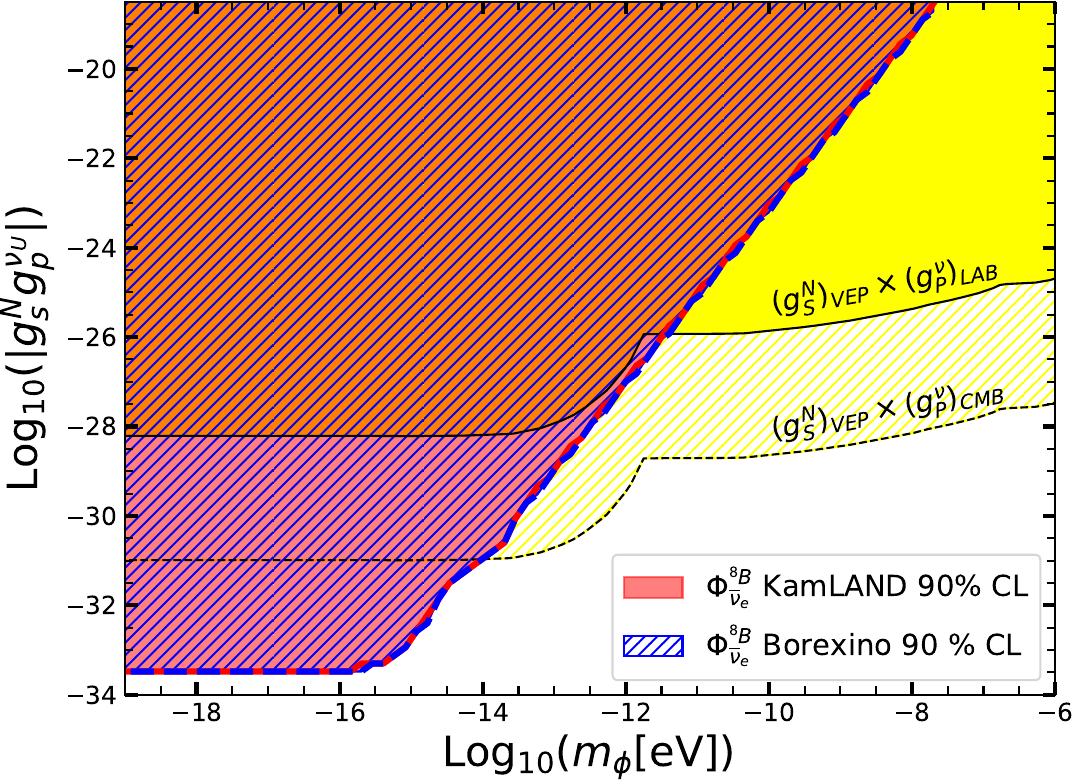}
  \caption{Left: Limits on $\bar{\nu}_e$ flux of astrophysical origin,
    as reported by KamLAND~\recite{KamLAND:2021gvi},
    Borexino~\recite{Borexino:2019wln} and
    Super-Kamiokande~\recite{Super-Kamiokande:2020frs,
      Super-Kamiokande:2021jaq, Super-Kamiokande:2023xup}.  For
    comparison, we show the expected solar $\bar{\nu}_e$ flux for
    several values of the model parameters as labeled in the figure.
    Right: Excluded region on the ($m_{\phi}$, $g_s^N g_p^{\nu_U}$)
    plane obtained for Majorana neutrinos obtained from the 90\% CL
    constraint reported by KamLAND~\recite{KamLAND:2021gvi} and
    Borexino~\recite{Borexino:2019wln}, Eqs.~\eqref{eq:anubound1} and
    \eqref{eq:anubound2} (see text for details).  For comparison we
    show as yellow region the bounds on the product of couplings from
    the product of the bound on $g_s^N$ from violation of weak
    equivalent principle (VEP) and the bounds on $g_p^\nu$ from either
    kinematic and rare effects in weak decays in laboratory
    experiments (LAB) or cosmology bounds from cosmic microwave
    background (CMB) (see Sec.~\ref{sec:disc}).}
  \label{fig:anures}
\end{figure}

Turning now to the bounds derived from the non observation of a flux
of solar antineutrinos, we plot in the left panel in
Fig.~\ref{fig:anures} a compilation of the model-independent limits on
$\bar{\nu}_e$ flux of astrophysical origin as reported by
KamLAND~\cite{KamLAND:2021gvi}, Borexino~\cite{Borexino:2019wln} and
Super-Kamiokande~\cite{Super-Kamiokande:2020frs,
  Super-Kamiokande:2021jaq, Super-Kamiokande:2023xup}.  In the same
panel we also show the predicted flux in the presence of the
helicity-flip potential for several model parameters.  From the figure
we see that the most stringent upper bounds on $\bar{\nu}_e$ come
KamLAND~\cite{KamLAND:2021gvi} and Borexino~\cite{Borexino:2019wln}
which constraint the $\bar{\nu}_e$ flux from the $\Nuc[8]{B}$ reaction
by
\begin{align}
  \label{eq:anubound1}
  &\text{KamLAND:}
  &\Phi^{\Nuc[8]{B}}_{\bar{\nu}_e} &< \hphantom{0}60~\text{cm}^{-2}\, \text{s}^{-1}
  ~\text{at 90\% CL for}~ E_\nu > 8.3~\text{MeV},
  \\
  \label{eq:anubound2}
  &\text{Borexino:}
  &\Phi^{\Nuc[8]{B}}_{\bar{\nu}_e} &< 138~\text{cm}^{-2}\, \text{s}^{-1}
  ~\text{at 90\% CL for}~ E_\nu > 7.8~\text{MeV}.
\end{align}
which the experiments translated into a bound on the corresponding
integrated oscillation probabilities
\begin{align}
  \nonumber
  P(\nu_e\to \overline{\nu}_e)
  &\equiv
  \frac{\int_{E_\text{thres}} \Phi_{\Nuc[8]{B}}(E_\nu)\, \sigma(E_\nu)\,
    P_{\nu_{eL} \to \overline{\nu}_{eR}}(E_\nu)\, dE_\nu}{\int_{E_\text{thres}}
    \Phi_{\Nuc[8]{B}}(E_\nu)\, \sigma(E_\nu)\, dE_\nu}
  \\[1mm]
  \label{eq:panubound}
  &\leq
  \begin{cases}
    4.1\times 10^{-5} &\text{for}~ E_\text{thres} = 8.3~\text{MeV} \\
    7.4\times 10^{-5} &\text{for}~ E_\text{thres} = 7.8~\text{MeV}
  \end{cases}
\end{align}
where we have used the normalization of the $\Phi_{\Nuc[8]{B}}(E_\nu)$
of the latest version of the SSM~\cite{B23Fluxes, Magg:2022rxb} which
results in the slight difference in the bounds on the probabilities in
Eq.~\eqref{eq:panubound} with respect to those quoted by the
experimental collaborations.

We show in the right panel of Fig.~\ref{fig:anures} the constraints on
the model parameters obtained by the projection of the bounds in
Eq.~\eqref{eq:panubound}.  For concreteness in this case we fix the
oscillation parameters to their best-fit values, however as already
noted the dependence of the bounds on the oscillation parameters is
very mild within their allowed ranges from global oscillation
analysis~\cite{Esteban:2020cvm, nufit-5.3}.  We show the results for
pseudoscalar coupling to $\nu_U$ but similar bounds apply to the other
cases (see Table~\ref{tab:bounds} for the results obtained with the
different flavour assumptions in Eq.~\eqref{eq:nucoup}).  As seen from
the figure the bounds imposed with KamLAND and Borexino constraints
are very similar.  In addition, we notice the expected mass
independent asymptotic behaviour for effectively infinite range
interactions when $m_\phi\lesssim 10^{-16}$, as well as the dependence
on $|g_s^N g_p^{\nu_U}| \,/\, m_\phi^2$ in the contact interaction
regime for $m_\phi\gtrsim 10^{-14}$~eV.

\section{Discussion}
\label{sec:disc}

In this work we have shown that solar neutrinos are a powerful probe
for interactions mediated by a spin-zero particle with pseudoscalar
couplings to neutrinos and scalar coupling to the nucleons in the
solar matter.  We have evaluated the neutrino spin-flavour preceding
potential generated by such interaction sourced by the nucleons in the
Sun's matter.  This SFP potential presents the distinct feature of
being most relevant for neutrinos with the lowest energies and
therefore can be robustly constrained by a global analysis of solar
oscillation results (in combination with KamLAND) in both cases of
Dirac or Majorana neutrinos.  In addition in the case of Majorana
neutrinos the SFP potential also generates a flux of solar
antineutrinos which is severely constrained from observations.
Quantitatively our results show that, because of the distinctive
energy dependence of the effect, the bounds obtained from the global
analysis of solar neutrino data for both Dirac or Majorana neutrinos
are comparable with those arising from the antineutrino constraints in
the Majorana case.

We have focused on the limit of ultra light mediators for which the
interaction length is comparable with the solar radius, finding that
the bounds obtained go beyond existing constraints on the relevant
couplings to nucleons and neutrinos.  In brief, couplings of scalars
to nucleons in this mass range are most strongly constrained by
experiments testing for violations of the weak equivalence principle
(VEP), in particular from the latest results by
MICROSCOPE~\cite{MICROSCOPE:2022doy} for masses below
$10^{-14}$~eV\footnote{In order to derive the bounds shown in
Figs.~\ref{fig:limitsnu} and~\ref{fig:anures} we have scaled the
exclusion region shown in Fig.~1 of Ref.~\cite{Berge:2017ovy}
corresponding to the first MICROSCOPE results~\cite{Touboul:2017grn}
with the latest MICROSCOPE data~\cite{MICROSCOPE:2022doy}.} and by
torsion balance experiments~\cite{Schlamminger:2007ht} (and also from
fifth force searches, see~\cite{Adelberger:2009zz} for a compilation)
in the range~$10^{-14}$~eV to~$10^{-6}$~eV.  In what respects the
pseudoscalar couplings to neutrinos in this mass range, they can be
tested in a variety of probes, from laboratory experiments to effects
in astrophysics and cosmology (for a recent review see
Ref.~\cite{Berryman:2022hds} and references therein).  Model
independent bounds on couplings to neutrinos in this mediator mass
range have been derived from kinematic and rate effects in meson,
charge lepton, Higgs and Z decays~\cite{Lessa:2007up,
  Pasquini:2015fjv, Berryman:2018ogk}: $g_p^{\nu_e}\lesssim 3\times
10^{-3}$, $g_p^{\nu_\mu}\lesssim 10^{-3}$, $g_p^{\nu_\tau}\lesssim
\text{few}\times 10^{-1}$, and from neutrinoless double beta decay
$g_p^{\nu_e}\lesssim 10^4$--$10^{-5}$~\cite{Berryman:2022hds}.
Stronger but more model dependent constraints can be derived from
effects in astrophysics (for example from supernova
cooling~\cite{Farzan:2002wx}) and cosmology.  In particular the latest
cosmic microwave background (CMB) data has been used to constraint
couplings to very light (effectively massless) pseudoscalars $g_p^\nu
\leq 7 \times 10^{-7}$~\cite{Forastieri:2019cuf}.  For sake of
comparison we show in Figs.~\ref{fig:limitsnu} and~\ref{fig:anures}
the range covered by the product of the bounds on $g_s^N$ from VEP
constraints and the bound on $g_p^\nu$ from laboratory (LAB) to CMB
constraints.  As seen in the figure the bounds derived in this work
overcome the existing bounds obtained with constraints from laboratory
experiments for mediator masses $m_\phi\lesssim 10^{-12}$~eV and with
those from CMB for $m_\phi\lesssim 10^{-14}$~eV.

We finish by commenting on the possible relevance of these constraints
for the proposed solution to the muon $g-2$ anomaly in
Ref.~\cite{Agrawal:2022wjm, Davoudiasl:2022gdg} which requires
$g_p^\mu g_s^N \gtrsim 10^{-30}$.  Taken at face values the
constraints derived in this work rule out the proposed solution in the
context of models in which the pseudoscalar coupling to the leptons of
a given generation are related by symmetry.  There is, however, a
caveat.  That conclusion is valid as long as the relation holds
between the coupling $g_p^{\nu_\mu}$ in Eq.~\eqref{eq:lagran} and a
muon coupling $g_p^\mu$ with interaction
\begin{equation}
  \label{eq:lagmu1}
  i\, g_p^\mu\,
  \phi\, \bar\mu \gamma^5\mu
\end{equation}
But pseudoscalar interactions for an on-shell fermion $\psi$ can
equally be cast as
\begin{equation}
  \label{eq:derivlag}
  \tilde{g}^\psi\, \frac{\partial_\mu\phi}{f}\,
  \overline{\psi}\, \gamma^\mu \gamma^5\, \psi \,,
\end{equation}
where $f$ sets the scale of the dimension-5 operator.  For muons the
relation between the couplings in Eq.~\eqref{eq:lagmu1}
and~\eqref{eq:derivlag} is $g_p^\mu = (2\, m_\mu / f)\,
\tilde{g}_p^\mu$.  For neutrinos the relation reads
\begin{equation}
\label{eq:Equality}
  \Big( U_\text{vac}^\dagger\,{g}_p^\nu\, U_\text{vac} \Big)_{ij}=
  \Big( U^\dagger_\text{vac}\, \tilde{g}_p^\nu\, U_\text{vac} \Big)_{ij}
  \frac{m_i+m_j}{f}
\end{equation}
where $m_i$ are the neutrino masses.  Thus if the symmetry relates the
neutrino and muon couplings $\tilde{g}_p^{\nu_\mu} \sim \tilde{g}^\mu$
the bounds here derived, $g_s^N g^{\nu_\mu} \lesssim 10^{-34}$, would
imply a weaker constraint, namely $g_s^N g^\mu \lesssim 10^{-34}\,
m_\mu \big/ m_{\nu_h} \sim 10^{-24}$ for the minimum possible value of
the heavier neutrino mass $m_{\nu_h}\sim 0.05$~eV as required from
oscillations.

\acknowledgments

We are thankful to Renata Zukanovich for pointing out the possible
connection between the proposed solution to the muon anomalous
magnetic moment and neutrinos which started this project and for her
collaboration in its early stages.  We are also grateful to Pilar
Coloma and Enrique Fernandez-Martinez for useful discussions during
the development of this work.
This project is funded by USA-NSF grant PHY-2210533 and by the
European Union's through the Horizon 2020 research and innovation
program (Marie Sk{\l}odowska-Curie grant agreement 860881-HIDDeN) and
the Horizon Europe research and innovation programme (Marie
Sk{\l}odowska-Curie Staff Exchange grant agreement
101086085-ASYMMETRY).  It also receives support from grants
PID2019-\allowbreak 105614GB-\allowbreak C21, PID2022-\allowbreak
126224NB-\allowbreak C21, PID2019-\allowbreak 110058GB-\allowbreak
C21, PID2022-\allowbreak 142545NB-\allowbreak C21, ``Unit of
Excellence Maria de Maeztu 2020-2023'' award to the ICC-UB
CEX2019-000918-M, grant IFT ``Centro de Excelencia Severo Ochoa''
CEX2020-001007-S funded by MCIN/AEI/\allowbreak 10.13039/\allowbreak
501100011033, as well as from grants 2021-SGR-249 (Generalitat de
Catalunya).  SA would like to thank the University of Barcelona for
its hospitality.  We also acknowledge the use of the IFT computing
facilities.

\appendix

\section{Derivation of the scalar-pseudoscalar potential}
\label{sec:appendix1}

Our starting point is the well-known relation between the static
potential generated by a non-relativistic source particle ---~here a
nucleon $f$ located at $\vec\rho$~--- felt by a probe particle
---~here a neutrino $\nu$ at position $\vec{x}$~--- to the elastic
scattering amplitude $\mathcal{M}$ of the process $\nu + f \to \nu + f
$ as computed in QFT in momentum space
\begin{equation}
  \label{eq:potential}
  V(\vec{r} \equiv \vec{x} - \vec\rho)
  = - \int \frac{d^3\vec{q}}{(2\pi)^3} \,
  e^{i\vec{q}\cdot\vec{r}}\, \mathcal{M}(\vec{q})
\end{equation}
with four-momentum transfer between the source particle and the
neutrino given by $q\equiv(0, \vec{q})\equiv p'-p$ where $p$ and $p'$
are the neutrino four-momenta before and after the elastic scattering,
which for a scalar-pseudoscalar interaction of a neutrino with mass
$m$
\begin{equation}
  \label{eq:lagran2}
  \mathcal{L} = g_s^f \phi \bar{f}f + ig_p^\nu \phi\bar{\nu}\gamma^5\nu
\end{equation}
reads
\begin{equation}
  \label{eq:amp}
  i\mathcal{M} = \frac{i(ig_s^f)(-g_p^\nu)}{q^2 - m_\phi^2}
  \Big[ \bar{u}_{r'}^f(k) u_{r}^f(k) \Big]
  \Big[ \bar{u}_{\lambda'}^\nu(p')\gamma^5 u_{\lambda}^\nu(p) \Big] \,.
\end{equation}
In writing Eq.~\eqref{eq:potential} we are implicitly assuming the
normalization of the spinors to be one particle per unit volume for
both nucleons and neutrinos.  Explicitly, in the Dirac representation
of the gamma matrices our choice for the spinor of fermion with
four-momentum $(E, \vec{p})$ and spin state $s$ is
\begin{equation}
  u_s(p) = \sqrt{\frac{E+m}{2E}}
  \begin{pmatrix}
    \chi_s
    \\
    \frac{\vec\sigma \cdot \vec{p}}{E+m} \chi_s
  \end{pmatrix}
\end{equation}
where $\chi_s$ are 2-spinors normalized as
$\chi^\dagger_{s'}\chi_s=\delta_{ss'}$.
With this choice, for the non-relativistic nucleons $\bar{u}_{r'}^f(k)
u_r^f(k) = \delta_{rr'}$, while for the neutrinos we can write
\begin{equation}
  \bar{u}_{\lambda'}^\nu(p')\gamma^5 u_{\lambda}^\nu(p)
  = \sqrt{\frac{E'+m}{2E'}}\sqrt{\frac{E+m}{2E}} \,
  \Big( \chi^{\dag}_{\lambda'} \vec{\sigma} \chi_{\lambda} \Big) \cdot
  \bigg( \frac{\vec{p}'}{E' + m} - \frac{\vec{p}}{E + m} \bigg) \,.
\end{equation}
Expanding to the lowest non-vanishing order in $\vec{q}$ we find
\begin{equation}
  \label{eq:nuline}
  \bar{u}_{\lambda'}^\nu(p')\gamma^5 u_{\lambda}^\nu(p) =
  \frac{\chi^{\dag}_{\lambda'} \vec{\sigma} \chi_{\lambda}}{2E} \cdot
  \Big[ \vec{q} - \Big(1 - \frac{m}{E} \Big) (\hat{n} \cdot \vec{q})\,
    \hat{n} \Big],
\end{equation}
where $\hat{n}$ denotes the neutrino direction.  Introducing
Eq.~\eqref{eq:nuline} in Eq.~\eqref{eq:amp} and evaluating the
$\vec{q}$ integral in Eq.~\eqref{eq:potential} one finds
\begin{equation}
  \label{eq:vpoint}
  V(\vec{r}) = - \frac{g_s^f g_p^\nu}{8\pi E}
  \bigg[ \vec{S}_{\lambda\lambda'} - \Big( 1 - \dfrac{m}{E} \Big)
    (\vec{S}_{\lambda\lambda'} \cdot \hat{n})\, \hat{n} \bigg] \cdot
  \vec{\nabla}_r \bigg(\frac{e^{-m_\phi r}}{r} \bigg)\,,
\end{equation}
where we have defined $\vec{S}_{\lambda\lambda'} \equiv
\chi^{\dag}_{\lambda'}\, \vec{\sigma}\, \chi_{\lambda}$ as a spin-like
vector quantity.  
In the non-relativistic limit $\vec{S}_{\lambda\lambda'}$ corresponds
to the spin of the neutrino, and Eq.~\eqref{eq:vpoint} reduces to the
well-known monopole-dipole potential~\cite{Moody:1984ba}
\begin{equation}
  V(r)
  = g_s^f g_p^\nu\, \frac{\hat\sigma_\nu \cdot \hat{r}}{8\pi\, m_\nu}\,
  \bigg[ \frac{m_\phi}{r} + \frac{1}{r^2} \bigg]\, e^{-m_\phi r}
\end{equation}
where $\hat\sigma_\nu$ is the direction of the neutrino spin.
Conversely for relativistic neutrinos one can identify $\lambda$ with
the helicity of the neutrino and using the explicit form of the
spinors:
\begin{equation}
  \vec{S}_{++} = -\vec{S}_{--} = \hat{n} \,,
  \qquad
  \vec{S}_{+-} = \vec{S}_{-+}^* = \hat{u} + i \hat{v} \,,
  \qquad
  (\text{with~} \hat{n} \perp \hat{u} \perp \hat{v})
\end{equation}
we get
\begin{equation}
  \label{eq:potpoint}
  V(\vec{r}) = -\frac{g_s^f\, g_p^\nu}{8\pi E}\,
  \bigg[ \delta_{\lambda\neq \lambda'} (\vec{\nabla}_r)_\perp
    \pm \delta_{\lambda\lambda'}\, \frac{m}{E} (\vec{\nabla}_r)_\parallel
    \bigg] \, \bigg( \frac{e^{-m_\phi r}}{r} \bigg)
\end{equation}
where
\begin{equation}
  (\vec{\nabla}_r)_\parallel \equiv \hat{n}\cdot \vec{\nabla}_r
  \quad\text{and}\quad
  (\vec{\nabla}_r)_\perp \equiv (\hat{u} + i\hat{v})\cdot \vec{\nabla}_r
\end{equation}
are the gradient operators along and perpendicular to the neutrino
direction respectively.  We notice that Eq.~\eqref{eq:potpoint}
displays the relativistic $1/\gamma = m/E$ factor suppression of the
helicity-conserving potential in analogy with the well-known results
for the potential generated by a magnetic field in the presence of a
neutrino magnetic moment~\cite{Akhmedov:1988hd, Akhmedov:1988kih}.
Conversely the scalar-pseudoscalar interaction produces a
helicity-flip potential which is proportional to the variation of the
potential in the direction perpendicular to that of the neutrino
propagation.

We are interested in taking the Sun as a source of the potential and
for that we integrate Eq.~\eqref{eq:potpoint} with the number density
of nucleon $f$ at position $\vec\rho$ to obtain the potential at
neutrino position $\vec{x}$ to be
\begin{equation}
  V(\vec{x}) =
  -\frac{g_s^f\, g_p^\nu}{8\pi\, E}\
  \bigg[ \delta_{\lambda\neq\lambda'} (\vec{\nabla}_x)_\perp
    \pm \delta_{\lambda\lambda'}\, \frac{m}{E} (\vec{\nabla}_x)_\parallel \bigg]
  \int N_f(\vec\rho)\,
    \frac{e^{-m_\phi |\vec\rho - \vec{x}|}}{|\vec\rho - \vec{x}|} \,
    d^3\vec\rho \,.
\end{equation}
where we have replaced $\vec{\nabla}_r $ with $-\vec{\nabla}_\rho =
\vec{\nabla}_x$ because the position of the neutrino is fixed in the
integral.

Since $N_f(\vec\rho)$ is spherically symmetric it is convenient to
choose $\hat{u}$ aligned with the plane containing the neutrino
trajectory and the center of the Sun, and $\hat{v}$ orthogonal to
it.  In this way $(\vec{\nabla}_x)_\perp$ will not generate any
imaginary part and consequently $V(\vec{x})$ will be real along the
entire trajectory.

\bibliographystyle{JHEPmod}
\bibliography{references}

\end{document}